\documentclass[prb,amsfonts,amssymb,showpacs,twocolumn,floatfix]{revtex4}
\usepackage{graphics,bm,natbib}
\begin{document}
\widetext
\title{Construction of Wannier functions from localized atomic-like orbitals}
\author{I. V. Solovyev}
\email[Electronic address: ]{solovyev.igor@nims.go.jp}
\affiliation{
Computational Materials Science Center (CMSC),\\
National Institute for Materials Science (NIMS),\\
1-2-1 Sengen, Tsukuba, Ibaraki 305-0047, Japan
}
\author{Z. V. Pchelkina}
\author{V. I. Anisimov}
\affiliation{
Institute of Metal Physics, Russian Academy of Sciences -- Ural Division,\\
620219 Ekaterinburg GSP-170, Russia
}
\date{\today}

\widetext

\begin{abstract}
The problem of construction of the Wannier functions (WFs) in a restricted
Hilbert space of eigenstates of the one-electron Hamiltonian
$\hat{H}$ (forming the so-called low-energy part of the
spectrum) can be formulated in several different ways. One
possibility is to use the projector-operator techniques, which pick
up a set of trial atomic orbitals and project them onto the given
Hilbert space. Another possibility is to employ the downfolding method,
which eliminates the high-energy part of the spectrum and
incorporates all related to it properties into the energy-dependence
of an effective Hamiltonian. We show that by modifying the
high-energy part of the spectrum of the original Hamiltonian
$\hat{H}$, which is rather irrelevant to the construction of
WFs in the low-energy part of the spectrum, these two
methods can be formulated in an absolutely exact and identical form,
so that the main difference between them is reduced to the choice of
the trial orbitals. Concerning the latter part of the problem, we argue that an
optimal choice for trial orbitals can be based on the maximization
of the site-diagonal part of the density matrix. This idea is
illustrated for a simple toy model, consisting of only two bands,
as well as for a more realistic example of $t_{2g}$ bands in V$_2$O$_3$.
Using the model analysis, we explicitly show that a bad choice of
trial orbitals can be linked to the \textit{discontinuity} of
phase of the Bloch waves in the reciprocal space, which leads to
the delocalization of WFs in the real space.
Nevertheless, such a discontinuity does not necessary contribute to
the matrix elements of $\hat{H}$ in the Wannier basis.
Similar tendencies are seen in realistic calculations for
V$_2$O$_3$, though with some variations caused by the
multi-orbital effects. An analogy with the search of the ground
state of a many-electron system is also discussed.
\end{abstract}



\maketitle


\onecolumngrid

\section{\label{sec:Intro}Introduction}

  The Wannier functions (WFs) play a key role in the solid-state physics, as they make a
direct connection between the reciprocal ${\bf k}$-momentum space and the real space
representations for the quantum-mechanical operators, which allows us to formulate the
problem
of electronic-structure calculations
for the infinite periodical systems in a basis of localized (Wannier) orbitals.\cite{Wannier}
The construction of WFs using first-principles electronic
structure calculations has attracted a considerable attention
recently.\cite{MarzariVanderbilt,WeiKu,Schnell,PRB04,Pavarini,Anisimov2005,Streltsov,Pavarini2,PRB06,condmat06}
The reason is
mainly twofold:
\begin{itemize}
\item[{(a)}]
WFs appear to be extremely useful for the visualization
and interpretation
of the behavior of the local quantities in the real space. For example,
in the electronic structure calculations based on the plane waves,
such a construction can add many new functionalities, which were initially
a privilege of
the methods working in the basis of localized atomic orbitals.\cite{MarzariVanderbilt,Nakamura}
\item[{(b)}]
Another important application of WFs is related with
the construction of the so-called ``\textit{ab initio} models for the
strongly-correlated materials'',
which are designed to deal with the
low-energy properties of these systems and go beyond the conventional
(and oversimplified in the case of the strongly-correlated systems)
local-density approximation (LDA).
In this case, the Wannier orbitals form the basis of the physical
Hilbert space, which is then used for the construction of the model (typically, Hubbard-type)
Hamiltonians.\cite{WeiKu,Schnell,PRB04,Pavarini,Anisimov2005,Streltsov,PRB06,condmat06}
\end{itemize}

  The choice of the Wannier function is not unique, and their
extension in the real space can vary depending on the procedure
and the parameters of calculations.
In order to illustrate the importance of this problem, one may remind
to the reader
that
one extreme example of WFs are the Bloch states,
which are totally delocalized in the real space. Of course, the
physics should not depend on whether the WFs are
localized or not, because all representations provide the complete
sets of the basis functions (for the low-energy part of the spectrum),
which are totally equivalent from the physical point of view.
However, from the viewpoint of practical applications, it is
more convenient to work with the localized orbitals. For example,
in the process of construction of the Hubbard Hamiltonian one would
always like
to keep only the site-diagonal elements of Coulomb and exchange
interactions and neglect all intersite interactions.
This is only possible when the Wannier basis is
sufficiently well localized in the real space, so that the overlap
between the functions centered at different lattice sites
is minimal.
The localization of the basis functions is also important in
applications dealing with the
response of the electronic system onto a perturbation
confined within a single atomic site,
like the orbital magnetization in solids.\cite{PRL05}

  Since the idea of construction of
``\textit{ab~initio} models for the strongly-correlated systems''
is still relatively
new,
there is no clear consensus among researchers
working in this area
about the
origin of apparent differences,
accuracy, and the scopes of applicability
of their results.\cite{PRB04,Pavarini,Streltsov,Pavarini2,PRB06,condmat06}
We believe that in such a situation it is very important to make a direct comparison
between different methods of calculation of WFs,
and prove several theorems regarding both the
formulation of the methods and the choice of the computational parameters, which directly control
the localization of WFs.
Particularly, we will prove rigorously that the downfolding method,
considered in Refs.~\onlinecite{PRB04}, \onlinecite{PRB06}, and \onlinecite{condmat06},
can be reformulated as a projector-operator method, employed in
Refs.~\onlinecite{WeiKu}, \onlinecite{Anisimov2005}, and \onlinecite{Streltsov}
(and, as an initial stage, in Ref.~\onlinecite{MarzariVanderbilt}).
This can be done by a simple
(scissor-operator-like) change of the full Hamiltonian of the system
in the high-energy part of the spectrum, which does
not contribute to the low-energy part.
This procedure has been already implemented in
Refs.~\onlinecite{PRB04}, \onlinecite{PRB06}, and \onlinecite{condmat06}.
Therefore, two methods are simply equivalent, and the main difference
comes from the form of trial orbitals, which are
used to generate the Wannier function in the projector-operator method
and, directly, the effective low-energy Hamiltonian in the downfolding method.
Concerning the last part of the problem, we will argue that
an optimal choice
for the trial wavefunctions
can be based on the maximization of the site-diagonal
part of the density matrix, as it was originally proposed in Ref.~\onlinecite{PRB04}.
We will also consider an analogy with the search of the ground state
of the many-electron systems.

\section{\label{sec:Method}Atomic orbitals and Wannier functions}

  We assume that there is certain set of orthonormalized atomic-like
orbitals $\{ \tilde{\chi}_{\bf R}^\alpha \}$ centered at the atomic sites $\{ {\bf R} \}$
and specified by the orbital indices $\{ \alpha \}$.
The basis is
complete and sufficient to reproduce the one-electron
band structure of a considered compound in the valent part of the spectrum.
The corresponding Hamiltonian matrix is denoted by $\hat{H}$.
The concrete examples of such bases could be the orthonormalized atomic
orbitals or the muffin-tin orbitals.\cite{LMTO}
We further assume that the basis orbitals are maximally localized
(in accordance with a certain criterion of maximal localization, the precise
form of which is not important here),\cite{MarzariVanderbilt}
in the sense that any linear combination of $\{ \tilde{\chi}_{\bf R}^\alpha \}$ will be either less
localized or have the same degree of the localization as $\{ \tilde{\chi}_{\bf R}^\alpha \}$.
Then, we immediately recognize that the basis functions $\{ \tilde{\chi}_{\bf R}^\alpha \}$,
selected in such a way, satisfy to all criteria of WFs,
and
can be regarded as
one of possible representations of
WFs for the full Hamiltonian $\hat{H}$.
Indeed, the basis functions $\{ \tilde{\chi}_{\bf R}^\alpha \}$ are localized, orthonormalized, and
(due to the property of completeness of the basis set)
any eigenvector of $\hat{H}$ in the valent part of the spectrum can be expressed as a
linear combination of $\{ \tilde{\chi}_{\bf R}^\alpha \}$.
This is a natural result and advantage of using the localized
(or atomic-like) basis.
In the plane-wave basis, the localized WFs can be constructed from
certain number of eigenvectors of $\hat{H}$ in the valent part of the spectrum,
for example, by minimizing the square of the position operator
$\langle {\bf r}^2 \rangle$.\cite{MarzariVanderbilt}
However, we would like emphasize that
this is nothing but an elegant way of constructing a compact atomic basis from the
extended
plane waves, a step which becomes rather unnecessary if one works
from the very beginning in
an appropriate basis of
atomic orbitals.
The precise criterion of the maximal localization (for example, the minimization of
$\langle {\bf r}^2 \rangle$) is not really important at this stage, because this is merely
a mathematical construction and depending on the considered physical property
one can introduce different criteria of the ``maximal localization''.

  However, what we typically need in the process of construction of the model Hamiltonians
for the strongly-correlated systems is different. For example,
for the solution of the many-electron problem
it is practically impossible to operate with a
large number of WFs of the full Hamiltonian $\hat{H}$.
Instead, one would like to concentrate on the evolution of a small number of bands,
located near the Fermi level, and construct the WFs only for this group
of bands. A concrete example is the $t_{2g}$ bands in many transition-metal oxides, which
are typically well separated from the rest of the spectrum.\cite{PRB04,PRB06}
Of course, the WFs for the $t_{2g}$ bands should be orthogonal to the
rest of the eigenstates of the full Hamiltonian $\hat{H}$. This causes an additional
complication and the basis functions $\{ \tilde{\chi}_{\bf R}^\alpha \}$,
though can be regarded as WFs for the full Hamiltonian $\hat{H}$,
are no longer the WFs for a limited subspace of
eigenstates of $\hat{H}$, restricted by the $t_{2g}$ bands.

  At present, there are two methods, which are commonly used for the construction of
WFs for a restricted number of states of the full Hamiltonian $\hat{H}$:
the projector-operator method
(for example, employed in Refs.~\onlinecite{MarzariVanderbilt}, \onlinecite{WeiKu},
\onlinecite{Anisimov2005}, and \onlinecite{Streltsov})
and the downfolding method
(employed in Refs.~\onlinecite{PRB04}, \onlinecite{PRB06}, and \onlinecite{condmat06}).

\subsection{\label{sec:projector}The Projector-Operator Method}

  In the projector-operator method,
each (nonorthonormalized) Wannier function is generated by
projecting
a trial basis function $| \tilde{\chi}_{{\bf R}}^t \rangle$
centered at the site ${\bf R}$
onto a chosen subset of bands
(for instance, the $t_{2g}$ bands, located near the Fermi level):
\begin{equation}
|W_{{\bf R}}^t \rangle = \hat{P} | \tilde{\chi}_{{\bf R}}^t \rangle,
\label{eqn:Wannier_definition}
\end{equation}
where
\begin{equation}
\hat{P} = \sum_{i \in t_{2g}} | \psi_i \rangle \langle \psi_i |
\label{eqn:projector}
\end{equation}
is the projector-operator onto the $t_{2g}$ bands, $\psi_i$ denotes the eigenstate
of $\hat{H}$, and $i$ is a joint index combining the band index and the position of the
momentum ${\bf k}$ in the first Brillouin zone.
The effective tight-binding Hamiltonian
$\hat{\mathfrak{h}}$$=$$\| \mathfrak{h}_{{\bf R}{\bf R}'}^{tt'} \|$
in the region of $t_{2g}$-bands is constructed from
the matrix elements of $\hat{H}$ in the basis of these
Wannier states:
\begin{equation}
\mathfrak{h}_{{\bf R}{\bf R}'}^{tt'} = \langle W_{\bf R}^t | \hat{H} | W_{{\bf R}'}^{t'} \rangle.
\label{eqn:HamiltonianP}
\end{equation}
The orbitals $\{ W_{{\bf R}}^t \}$ shall be
further orthonormalized:
\begin{equation}
| \tilde{W}_{\bf R}^t \rangle = \sum_{{\bf R}'t'} | W_{{\bf R}'}^{t'} \rangle
[ \hat{\cal S}^{-1/2} ]_{{\bf R}'{\bf R}}^{t't},
\label{eqn:orthonormalization}
\end{equation}
where
$\hat{\cal S}$$=$$\| {\cal S}_{{\bf R}'{\bf R}}^{t't} \|$
is the overlap matrix:
\begin{equation}
{\cal S}_{{\bf R}'{\bf R}}^{t't} = \langle W_{{\bf R}'}^{t'} | W_{\bf R}^t \rangle
\equiv \langle \tilde{\chi}_{{\bf R}'}^{t'} | \hat{P} | \tilde{\chi}_{{\bf R}}^{t} \rangle .
\label{eqn:overlapP}
\end{equation}
Then, the effective tight-binding Hamiltonian $\hat{\tilde{\mathfrak{h}}}$ in the basis of
orthonormalized Wannier orbitals $\{ \tilde{W}_{{\bf R}}^t \}$ takes the following form:
\begin{equation}
\hat{\tilde{\mathfrak{h}}} = \hat{\cal S}^{-1/2} \hat{\mathfrak{h}} \hat{\cal S}^{-1/2}.
\label{eqn:TBP}
\end{equation}
$\hat{\tilde{\mathfrak{h}}}$ is typically regarded as the kinetic-energy part
of an effective Hubbard-type model in the projector-operator method.\cite{WeiKu,Anisimov2005,Streltsov}

  For the subsequent discussions, it will be also convenient to
write the overlap matrix $\hat{\cal S}$
in the form: $\hat{\cal S}$$=$$\hat{P}^{tt}$, where $\hat{P}^{tt}$ is the block
of matrix elements of the projector operator (\ref{eqn:projector}) in the basis of
trial orbitals $\{ \tilde{\chi}_{{\bf R}}^t \}$.
Since $\hat{P}$ commutes with $\hat{H}$, the projector-operator method guarantees that
$\hat{\tilde{\mathfrak{h}}}$ has the same set of eigenvalues in the region of $t_{2g}$-bands as
the original Hamiltonian $\hat{H}$.

\subsection{\label{sec:downfolding}The Downfolding Method}

  The conventional downfolding method also implies that all atomic basis functions
can be
divided into two parts:
$\{ \tilde{\chi}_{\bf R} \}$$=$$\{ \tilde{\chi}_{\bf R}^t \}$$\oplus$$\{ \tilde{\chi}_{\bf R}^r \}$,
so that the low-energy part of the spectrum in the direct proximity to the Fermi
level is mainly composed of the $\{ \tilde{\chi}_{\bf R}^t \}$-states, while $\{ \tilde{\chi}_{\bf R}^r \}$
is the rest of the basis states, which mainly contribute to the high-energy part.
Then, each eigenstate $| \psi_i \rangle$ of the full Hamiltonian $\hat{H}$
can be presented identically as the sum
$|\psi_i \rangle$$=$$|\psi_i^t \rangle$$+$$|\psi_i^r \rangle$,
where $|\psi_i^t \rangle$ and $|\psi_i^r \rangle$
are expanded over the basis states of correspondingly ``$t$''- and ``$r$''-type,
and the matrix equation for $| \psi_i \rangle$ takes the form:
\begin{eqnarray}
( \hat{H}^{tt}-\omega ) | \psi_i^t \rangle  +  \hat{H}^{tr} | \psi_i^r \rangle & = & 0  \label{eqn:seceq1}\\
\hat{H}^{rt} | \psi_i^t \rangle  +  ( \hat{H}^{rr}-\omega ) | \psi_i^r \rangle & = & 0, \label{eqn:seceq2}
\end{eqnarray}
where $\hat{H}^{t(r)t(r)}$ are the blocks of
matrix elements of $\hat{H}$ in the basis of ``$t$''(``$r$'')-states.
Then, $| \psi_i^r \rangle$ can be expressed from Eq.~(\ref{eqn:seceq2}) as
\begin{equation}
| \psi_i^r \rangle = - ( \hat{H}^{rr}-\omega )^{-1} \hat{H}^{rt} | \psi_i^t \rangle
\label{eqn:relimination}
\end{equation}
and substituted into Eq.~(\ref{eqn:seceq1}).
This yields an effective $\omega$-dependent Hamiltonian, which acts
formally only on $| \psi_i^t \rangle$:
\begin{equation}
\hat{h}(\omega) = (\hat{H}^{tt} - \omega) - \hat{H}^{tr}
(\hat{H}^{rr} - \omega)^{-1}\hat{H}^{rt}.
\label{eqn:Heff}
\end{equation}
However, $| \psi_i^t \rangle$ is only a part of the eigenvector, which is not orthonormalized.
Therefore,
$\hat{h}(\omega)$ should be additionally transformed to an orthonormal
basis:
\begin{equation}
\hat{\tilde{h}}(\omega) = \hat{S}^{-1/2}(\omega) \hat{h}(\omega) \hat{S}^{-1/2}(\omega) + \omega.
\label{eqn:TB}
\end{equation}
This transformation is specified by the overlap matrix
\begin{equation}
\hat{S}(\omega)=1+\hat{H}^{tr}
(\hat{H}^{rr}-\omega)^{-2}\hat{H}^{rt},
\label{eqn:overlapDF}
\end{equation}
which is obtained after the substitution of
$| \psi_i^r \rangle$, given by Eq.~(\ref{eqn:relimination}), into
the normalization condition for the full eigenvector
$| \psi_i \rangle$: $\langle \psi_i^t | \psi_i^t \rangle$$+$$\langle \psi_i^r | \psi_i^r \rangle$$=$$1$.
$\hat{\tilde{h}}(\omega_0)$, which is typically calculated in the center of gravity
of the low-energy bands ($\omega_0$)
constitutes the kinetic-energy part of the effective Hubbard-type model
in the downfolding method.\cite{PRB04,PRB06,condmat06}

  Although the downfolding method does not explicitly require
the construction of WFs, they are certainly implied
also in this approach and, at least, can be formally reconstructed from
$\hat{\tilde{h}}(\omega_0)$.\cite{PRB06}

\subsection{\label{sec:downfoldingasprojector}Downfolding as a Projector-Operator Method}

  The conventional downfolding method is exact. However, this property is enforced by
the $\omega$-dependence of $\hat{\tilde{h}}$, which
is hardly useful from the
practical point of view.
Formally, for each $\psi_i$, $\omega$ in Eq.~(\ref{eqn:TB}) should coincide with the
eigenvalue of $\hat{H}$ corresponding to this $\psi_i$.
Moreover, $\hat{\tilde{h}}(\omega)$ retains an excessive information about $\hat{H}$,
and the full spectrum of
eigenvalues and eigenfunctions of the original Hamiltonian $\hat{H}$
can be formally derived from $\hat{\tilde{h}}(\omega)$.
However, typically we do not need such a redundant information and
would like to use
$\hat{\tilde{h}}$ only in order to describe a small group of electronic states
located near the Fermi level, and do it in the most exact form.

  Therefore, what we want to do next is to show that
in a restricted subspace of eigenstates $\{  \psi_i \}$ of
the original Hamiltonian $\hat{H}$, the downfolding method can be
reformulated as a projector-operator method and retains all
attractive features of the latter:
namely, it becomes exact
and does not depend on $\omega$.
The trick
(which was actually implemented in Refs.~\onlinecite{PRB04} and \onlinecite{PRB06})
is to replace
the original Hamiltonian $\hat{H}$
in the downfolding method
by certain modified Hamiltonian $\hat{\cal H}$, which has the same set of eigenvalues
$\{ \varepsilon_i \}$ and eigenfunctions $\{ \psi_i \}$ in the region of $t_{2g}$-bands.
The rest of the eigenstates is not important for our purposes and can be shifted from the
valent part of the spectrum to the region of infinite energies, specified by the parameter $\epsilon$.
Hence, the modified Hamiltonian $\hat{\cal H}$ can be taken in the form:
\begin{equation}
\hat{\cal H} = \sum_{i \in t_{2g}} | \psi_i \rangle \varepsilon_i \langle \psi_i |
+ \epsilon \hat{P}_\perp \equiv \hat{P} \hat{H} \hat{P} + \epsilon \hat{P}_\perp,
\label{eqn:modifiedH}
\end{equation}
where $\hat{P}_\perp$$=$$\hat{1}$$-$$\hat{P}$ is the projector operator to the subspace orthogonal to the
$t_{2g}$-bands.
According to the choice of the
$\{ \tilde{\chi}_{\bf R}^t \}$- and
$\{ \tilde{\chi}_{\bf R}^r \}$-basis functions
in the downfolding method, the latter
mainly contribute to the high-energy part of the spectrum,
and the overlap between $\psi_i$ and any linear combination of $\{ \tilde{\chi}_{\bf R}^r \}$
should be small for the low-energy bands. Therefore, all eigenvalues of $\hat{\cal H}^{rr}$
are of the order of $\epsilon$ and located in the high-energy part of the spectrum.
Then, it is intuitively clear that,
for $\epsilon$$\to$$\infty$,
the $\omega$-dependence in Eq.~(\ref{eqn:relimination})
can be neglected.
Therefore,
the method should be exact and not depend on $\omega$.
In order to prove it rigorously, it is convenient to use the idempotency
of the projector operator: $\hat{P}_\perp^2$$=$$\hat{P}_\perp$. This yields the following identities
for the matrix elements $\hat{P}^{t(r)t(r)}_\perp$ in the basis of ``$t$''(``$r$'')-states:
\begin{equation}
\hat{P}_\perp^{tt}\hat{P}_\perp^{tt} + \hat{P}_\perp^{tr}\hat{P}_\perp^{rt} = \hat{P}_\perp^{tt},
\label{eqn:Palgebra1}
\end{equation}
\begin{equation}
\hat{P}_\perp^{tt}\hat{P}_\perp^{tr} + \hat{P}_\perp^{tr}\hat{P}_\perp^{rr} = \hat{P}_\perp^{tr},
\label{eqn:Palgebra2}
\end{equation}
and
\begin{equation}
\hat{P}_\perp^{rt}\hat{P}_\perp^{tt} + \hat{P}_\perp^{rr}\hat{P}_\perp^{rt} = \hat{P}_\perp^{rt}.
\label{eqn:Palgebra3}
\end{equation}
Then, using Eqs.~(\ref{eqn:Palgebra1}), (\ref{eqn:Palgebra2}), and (\ref{eqn:Palgebra3}), one can prove the
following identities, which are valid for the modified Hamiltonian (\ref{eqn:modifiedH})
in the limit $\epsilon$$\to$$\infty$:
\begin{enumerate}
\item
The overlap matrix (\ref{eqn:overlapDF})
becomes
$$
\hat{S} = ( \hat{1} - \hat{P}_\perp^{tt} )^{-1} = \hat{\cal S}^{-1},
$$
which is the inverse overlap matrix of the projector-operator method (\ref{eqn:overlapP});
\item
$\hat{P}_\perp^{tt} = \hat{P}_\perp^{tr} (\hat{P}_\perp^{rr})^{-1} \hat{P}_\perp^{rt}$.
Therefore, all terms proportional to $\epsilon$ in the Hamiltonian (\ref{eqn:Heff})
are exactly cancelled out;
\item
The $\omega$-dependent part of the Hamiltonian (\ref{eqn:Heff}) can be transformed to
the form: $-$$\omega \hat{\cal S}^{-1}$;
\item
The remaining part of the Hamiltonian (\ref{eqn:Heff}), which depends neither on $\epsilon$ nor
on $\omega$ takes the following form:
$\hat{\mathfrak{h}}^{tt}$$-$$\hat{\mathfrak{h}}^{tr} (\hat{P}_\perp^{rr})^{-1} \hat{P}_\perp^{rt}$$-$$
\hat{P}_\perp^{tr} (\hat{P}_\perp^{rr})^{-1} \hat{\mathfrak{h}}^{rt}$$+$$
\hat{P}_\perp^{tr} (\hat{P}_\perp^{rr})^{-1} \hat{\mathfrak{h}}^{rr} (\hat{P}_\perp^{rr})^{-1} \hat{P}_\perp^{rt}$,
where $\hat{\mathfrak{h}}^{t(r)t(r)}$ stand for the matrix elements of the operator $\hat{P}\hat{H}\hat{P}$ in the
basis of ``$t$''(``$r$'')-states.
\end{enumerate}
Moreover,
it is easy to see that
$\hat{\mathfrak{h}}^{tt}$ coincides with the Hamiltonian matrix
(\ref{eqn:HamiltonianP}) of the projector-operator method prior the orthonormalization.
Then, by replacing one of
the projector operators $\hat{P}$ in
$\hat{P} \hat{H} \hat{P}$
by $\hat{P}^2$ and writing explicitly the product of
two operators in the basis of ``$t$''(``$r$'')-states, one obtains the following identities:
$$
\hat{\mathfrak{h}}^{rr} =
\hat{P}^{rr} \hat{\mathfrak{h}}^{rr} + \hat{P}^{rt} \hat{\mathfrak{h}}^{tr},
$$
$$
\hat{\mathfrak{h}}^{rt} =
\hat{P}^{rt} \hat{\mathfrak{h}}^{tt} + \hat{P}^{rr} \hat{\mathfrak{h}}^{rt},
$$
and
$$
\hat{\mathfrak{h}}^{tr} = \hat{\mathfrak{h}}^{tr} \hat{P}^{rr} + \hat{\mathfrak{h}}^{tt} \hat{P}^{tr}.
$$
Using these properties, it is straightforward to verify that
the Hamiltonian matrix (\ref{eqn:Heff}) of the downfolding method takes the
following form:
$$
\hat{h}(\omega) = {\cal S}^{-1} \hat{\mathfrak{h}}^{tt} {\cal S}^{-1} - \omega {\cal S}^{-1}.
$$
Then,
the orthonormalization transformation (\ref{eqn:TB}) yields
the following tight-binding Hamiltonian:
$$
\hat{\tilde{h}} = {\cal S}^{-1/2} \hat{\mathfrak{h}}^{tt} {\cal S}^{-1/2},
$$
which is totally equivalent to the tight-binding Hamiltonian (\ref{eqn:TBP}) obtained in the
projector-operator method.

   Thus, we have proven that by introducing the modified Hamiltonian of the form
(\ref{eqn:modifiedH}), the downfolding method can be naturally reformulated as
the projector-operator method. Therefore,
as long as we are interested in the low-energy properties
of the system, these two methods are equivalent.

  Here, one comment is in order. In our proof, we have used the fact
that the overlap between
$\{ \tilde{\chi}_{\bf R}^r \}$ orbitals and the eigenstates $\{ \psi_i \}$
of the Hamiltonian $\hat{H}$ should be small for the low-energy bands.
However, this implies a specific choice for the orbitals
$\{ \tilde{\chi}_{\bf R}^r \}$ (and also for $\{ \tilde{\chi}_{\bf R}^t \}$),
for which the equivalence between the downfolding and the projector-operator
methods actually takes place. If this property is not satisfied, and the
overlap between $\{ \tilde{\chi}_{\bf R}^r \}$ and $\{ \psi_i \}$ is large,
the downfolding method will eventually breaks down, as we will clearly see it
in our calculations for V$_2$O$_3$ below.

\subsection{\label{sec:analogy}An Analogy with the Many-Electron Problem}

  The equivalence between downfolding and
projector-operator methods appears to be more generic and has a direct implication to the theory
of many-electron systems. Indeed, any many-electron state $\Psi_i$
can be expanded over the basis of Slater determinants $\{ \Phi \}$, which play
the same role as atomic orbitals in
the one-electron version of the downfolding method.
Suppose that the many-electron ground state is nondegenerate and
one can identify a
single Slater determinant ($\Phi_G$),
which has the largest weight in the many-electron wavefunction $| \Psi_G \rangle$
corresponding to the ground state. In practice, such $ | \Phi_G \rangle$ can be obtained
in the frameworks of one-electron Hartree-Fock (or any other) theory.
Then, each $\Psi_i$ can be formally presented in the form
$| \Psi_i \rangle$$=$$| \Psi_i^t \rangle$$+$$| \Psi_i^r \rangle$,
where $| \Psi_i^t \rangle$ is the part proportional
to $| \Phi_G \rangle$ and $| \Psi_i^r \rangle$ is expanded over the rest of the basis
of Slater determinants.
$| \Psi_i^r \rangle $ can be further eliminated by applying
the downfolding method
to the true many-electron Hamiltonian $\hat{\mathbb{H}}$
and its eigenfunctions $\{ \Psi_i \}$. By doing so, the solution of the
many-electron problem can be reformulated in terms of
an $\omega$-dependent self-energy $\hat{\Sigma}(\omega)$,\cite{AGD}
in a restricted Hilbert space of many-electron states,
which is formally spanned by only one Slater determinant
$| \Phi_G \rangle $.
Nevertheless,
such a self-energy contains all information about the ground state as well as the
excited states of the many-electron system, which is formally incorporated
into the $\omega$-dependence of $\hat{\Sigma}$.

  However, the analysis presented
in our work clearly shows that if we are interested only in
the ground-state properties of the many-electron system, the problem
can be reformulated in terms of the projection onto the true
ground state
$\hat{P}_G$$=$$| \Psi_G \rangle \langle \Psi_G |$.
In this case, the Slater determinant $| \Phi_G \rangle$ plays
the role of the trial wavefunction, which makes a formal correspondence
between the true many-electron ground state $| \Psi_G \rangle$ and
an auxiliary state of independent quasiparticles, represented by $| \Phi_G \rangle$.
As long as we are concerned with the ground-state properties, such a formulation
is exact and does not depend on $\omega$.
Of course, the direct application of the projector-operator method to the
many-electron problem is hardly meaningful from the practical point of view,
because it implies the knowledge of the true many-electron ground state
$| \Psi_G \rangle$ at the first place. However,
this simple example,
suggested by the projector-operator method,
clearly shows that, in principle,
the procedure of obtaining
the total energy of a many-electron system
can be formulated:
\begin{itemize}
\item[{(a)}]
in a restricted Hilbert space, which is
spanned by only one Slater determinant;
\item[{(b)}]
in terms of an $\omega$-independent self-energy (or an effective potential).
\end{itemize}
This is a simple but exact property of the
total energy of a many-electron system, which does not relies on
any approximations.
It reminiscences some basic theorems of the density-functional theory (DFT),\cite{KohnSham}
and provides some insight into why
the same many-electron problem can be formulated in two
seemingly different ways, one of which is
DFT and the other one is the self-energy approach.

\section{\label{sec:trial}Choice of trial orbitals}

   Now let us come back to the construction of WFs and
specify the choice of
the trial orbitals $\{ \tilde{\chi}_{\bf R}^t \}$.

  As it was already pointed out in the introduction, the basis functions $\{ \tilde{\chi}_{\bf R}^\alpha \}$
can be regarded as WFs of the full Hamiltonian $\hat{H}$.
Each basis function is localized around a central atomic site and
satisfy certain criterion of the ``maximal localization'' such that any
linear combination of $\{ \tilde{\chi}_{\bf R}^\alpha \}$ will be ``less localized''
or at least have the same degree of the localization as the original basis set $\{ \tilde{\chi}_{\bf R}^\alpha \}$.
However,
this is not necessarily true if one wants to construct the WFs
only for some part of the electronic structure, specified by certain
(restricted)
set of eigenstates $\{ \psi_i \}$ of the Hamiltonian $\hat{H}$.
Due to the additional orthogonality condition to other bands, such a Wannier function
will inevitably be a linear combination of $\{ \tilde{\chi}_{\bf R}^\alpha \}$ and, hence,
a less localized function in comparison with the trial atomic
orbital $\tilde{\chi}_{\bf R}^t$.
Nevertheless, one can formulate the problem in a slightly different way and
ask which atomic orbital centered at the single atomic site
is the best representation for the Wannier orbital. Therefore, we search a new
set of orthonormalized trial
orbitals in the form:
\begin{equation}
| \tilde{\phi}^t_{\bf R} \rangle = \sum_{\alpha} c_{\bf R}^{\alpha} | \tilde{\chi}_{\bf R}^{\alpha} \rangle,
\label{eqn:newtrial}
\end{equation}
and find the coefficients $\{ c_{\bf R}^{\alpha} \}$ from
the condition that maximizes the
projections $\langle \tilde{\phi}^t_{\bf R} | W^t_{\bf R} [\tilde{\phi}^t_{\bf R}] \rangle$,
where $|W^t_{\bf R} [\tilde{\phi}^t_{\bf R}]\rangle$ is the nonorthonormalized Wannier function
constructed from $| \tilde{\phi}^t_{\bf R} \rangle$ using the projector-operator method:
$|W^t_{\bf R} [\tilde{\phi}^t_{\bf R}]\rangle$$=$$\hat{P} | \tilde{\phi}^t_{\bf R} \rangle$.
It will automatically guarantee that $| \tilde{\phi}^t_{\bf R} \rangle$ is the best
single-orbital representation for the Wannier function in the projector-operator method
among the trial orbitals of the form (\ref{eqn:newtrial}). By substituting
$|W^t_{\bf R} [\tilde{\phi}^t_{\bf R}]\rangle$ into expression for the projection
$\langle \tilde{\phi}^t_{\bf R} | W^t_{\bf R} [\tilde{\phi}^t_{\bf R}] \rangle$, one can
easily find that the problem is reduced to the maximization of the functional
$$
D = \max_{ \{ c_{\bf R}^{\alpha} \} } \left\{ \langle \tilde{\phi}^t_{\bf R} | \hat{P} | \tilde{\phi}^t_{\bf R} \rangle -
\lambda ( \langle \tilde{\phi}^t_{\bf R} | \tilde{\phi}^t_{\bf R} \rangle - 1 ) \right\}
$$
with respect to the coefficient $\{ c_{\bf R}^{\alpha} \}$, where the Lagrange multipliers $\{ \lambda \}$
enforce the orthonormalization condition for the new trial orbitals (\ref{eqn:newtrial}).
Then, the maximization of $D$ is equivalent to the diagonalization of
$\hat{P}_{\bf RR}$$=$$\| \langle \tilde{\chi}_{\bf R}^{\alpha} | \hat{P} |
\tilde{\chi}_{\bf R}^{\alpha'} \rangle \|$, which is nothing but the site-diagonal part of the
density matrix constructed for the $t_{2g}$ bands in the basis of atomic orbitals
$\{ \tilde{\chi}_{\bf R}^{\alpha} \}$. After the diagonalization,
we should simply pick up $n$ eigenstates
$ \{ \tilde{\phi}^t_{\bf R} \}$,
corresponding to the $n$ maximal eigenvalues $\{ \lambda \}$, where $n$ is the number
of Wannier orbitals centered at the atomic site ${\bf R}$.\cite{comment.1}
These $ \{ \tilde{\phi}^t_{\bf R} \}$ will automatically maximize $D$.
This procedure has been proposed in Ref.~\onlinecite{PRB04} without proof. Then, some
intuitive arguments in support of the diagonalization of the
density matrix have been given in Ref.~\onlinecite{PRB06}.
Here, we have argued that it
can be derived on the basis of a rigorous variational principle.

\section{\label{sec:results}Results and Discussions}

\subsection{\label{sec:toy}Toy Model}

  The simplest model, which nicely illustrates
the main idea of the previous Section
is the one-dimensional chain of atoms $\{ R \}$
(though the spacial dimensionality is not really important here
because a similar conclusion can be obtained in
two and three dimensions).
We assume that the atomic basis at each
site consists of two orthonormal orbitals, say $| \tilde{\chi}_R^1 \rangle$
and $| \tilde{\chi}_R^2 \rangle$.
The full Hamiltonian $\hat{H}$ is specified by two parameters:
the splitting $\Delta$ between atomic levels $1$ and $2$,
and the nearest-neighbor transfer integral $t$ ($<$$0$), which is assumed to
be the same for all pairs of orbitals,
$$
\hat{H}_{RR'} =
\left(
\begin{array}{cc}
0 &      0 \\
0 & \Delta \\
\end{array}
\right) \delta_{R,R'}
+
\left(
\begin{array}{cc}
t & t \\
t & t \\
\end{array}
\right) (\delta_{R,R'+1} + \delta_{R,R'-1}).
$$
The Hamiltonian can be easily diagonalized in the reciprocal ($k$) space.
This yields two eigenstates: $| \psi_k^- \rangle$ and $| \psi_k^+ \rangle$.
Our goal is to construct the WFs for the lowest eigenstate $| \psi_k^- \rangle$,
which can be presented as
$$
| \psi_k^- \rangle = \cos \theta_k | \tilde{\chi}_k^1 \rangle +
\sin \theta_k | \tilde{\chi}_k^2 \rangle,
$$
where $\tilde{\chi}_k^1$ and $\tilde{\chi}_k^2$ are the Fourier transforms of
$\{ \tilde{\chi}_R^1 \}$ and $\{ \tilde{\chi}_R^2 \}$, respectively,
and $\cos \theta_k$ is nonnegative (that is defined by a proper choice of the phase of $| \psi_k^- \rangle$).
If $\Delta$ is large, this eigenstate will be composed mainly by the atomic orbitals of the
first type, i.e. $\{ \tilde{\chi}_R^1 \}$, which correspond
to the largest diagonal matrix elements of the density matrix. Therefore,
according to the procedure considered in the previous Section,
in order to obtain the maximally localized representation for the WFs,
$\{ \tilde{\chi}_R^1 \}$ should be used as the trial orbitals.

  Now, imagine that we do not know this property and continue to use for the trial
orbital $| \tilde{\chi}_R^t \rangle$ a linear combination of
$| \tilde{\chi}_R^1 \rangle$ and $| \tilde{\chi}_R^2 \rangle$:
$$
| \tilde{\chi}_R^t \rangle = \cos \beta | \tilde{\chi}_R^1 \rangle +
                                   \sin \beta | \tilde{\chi}_R^2 \rangle.
$$
Then, what will happen?

\begin{figure}[h!]
\begin{center}
\resizebox{12cm}{!}{\includegraphics{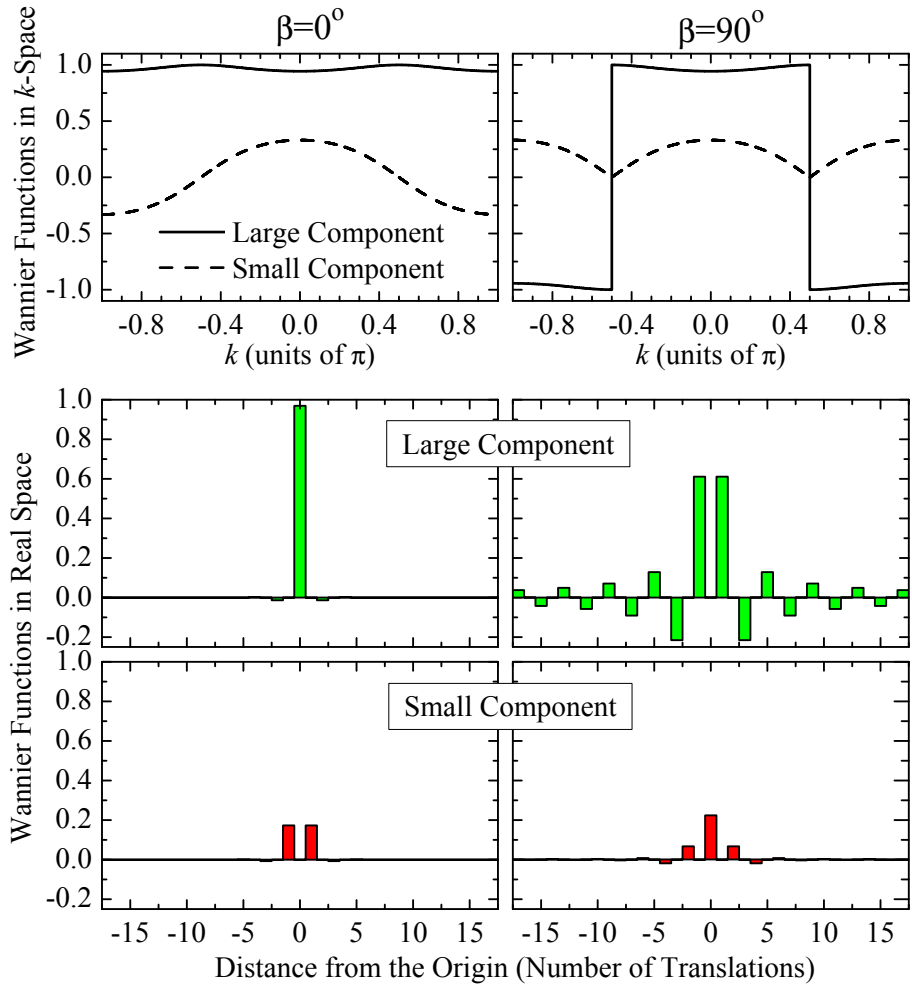}}
\end{center}
\caption{\label{fig.WFmodel}
(Color online)
Results of model calculations for $t/\Delta$$=$$-$$0.2$:
Wannier functions corresponding to the
trial orbitals
of the form
$| \tilde{\chi}_R^t \rangle$$=$$\cos \beta |\tilde{\chi}_R^1\rangle$$+
$$\sin \beta |\tilde{\chi}_R^2\rangle$
with $\beta$$=$$0^\circ$ (left) and $\beta$$=$$90^\circ$ (right).
The upper panel shows the behavior in the reciprocal ($k$) space, and the lower
panel shows the distance-dependence of Wannier functions after the
Fourier transformation to the real space. The `Large Component' is the
projection onto the atomic orbital $| \tilde{\chi}_R^1 \rangle$, and the
`Small Component' is the projection onto the atomic orbital $| \tilde{\chi}_R^2 \rangle$
(see text for details).
}
\end{figure}

  It is straightforward to verify that the Wannier function
$| \tilde{W}_k^- \rangle$ (in the reciprocal space) obtained after the
projection of $| \tilde{\chi}_R^1 \rangle$
onto $| \psi_k^- \rangle \langle \psi_k^- |$ and the orthonormalization
(\ref{eqn:orthonormalization}) takes the following form
$$
| \tilde{W}_k^- \rangle = {\rm sgn} [\cos(\theta_k-\beta)] | \psi_k^- \rangle,
$$
where $\rm{sgn}[ ... ]$ stands for the sign of the argument.
This means that $| \tilde{W}_k^- \rangle$ differs from $| \psi_k^- \rangle$
by a phase, which is controlled by $\beta$ and depends on $k$.
For small $\beta$, ${\rm sgn} [\cos(\theta_k$$-$$\beta)]$$=$$1$ for all $k$.
Therefore, $| \tilde{W}_k^- \rangle$ is a smooth function in the reciprocal space
and its Fourier image is
well localized in the real space (Fig.~\ref{fig.WFmodel}).
However, when the angle $\beta$ increases (and starting from certain critical value of $\beta$),
${\rm sgn} [\cos(\theta_k$$-$$\beta)]$
becomes a \textit{discontinuous function} of $k$.
The position of this discontinuity depends both on $\beta$ and on the ratio $t/\Delta$.
Certainly, the discontinuity will affect the spread of
WFs and make them much less localized in the real space.
Thus,
the model nicely illustrates the connection of the
localization of the WFs
in the real space with the problem of phase
of the Bloch waves in the reciprocal space.

\begin{figure}[h!]
\begin{center}
\resizebox{12cm}{!}{\includegraphics{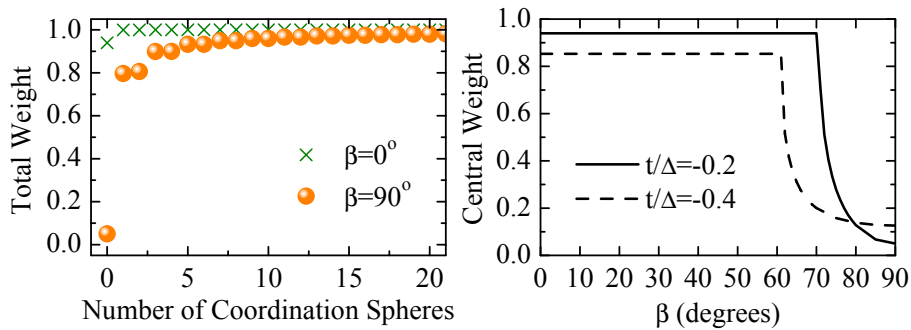}}
\end{center}
\caption{\label{fig.localizationmodel}
(Color online)
Left panel shows the convergence of the total weight of the Wannier
function in the real space for two types of
trial orbitals (specified by the parameter $\beta$).
Right panel shows the weight of the Wannier function at the central site
for two sets of parameters of the model Hamiltonian versus the angle $\beta$.
}
\end{figure}

  The degree of localization of the WFs in the real space
as a function of $\beta$ is explained in Fig.~\ref{fig.localizationmodel}.
For $\beta$$=$$0$, when the trial orbital corresponds to the largest
eigenvalue of the local density matrix, the Wannier function spreads
only over the central and nearest-neighbor atomic sites (which
for $t/\Delta$$=$$-$$0.2$
accumulate $99.9$\% of the total weight of the Wannier function).
When the angle $\beta$ increases, the weight of the Wannier function transfers
(abruptly, starting from certain value of $\beta$) from the central
part to more remote atomic sites. Hence, the Wannier function becomes
less localized. This is clearly manifested in the
extremely
slow convergence of the total weight
of the Wannier function
in the real space
(for example, for
$t/\Delta$$=$$-$$0.2$ and $\beta$$=$$90^\circ$,
$99.9$\% of the total weight can be regained only after
including about 200 coordination spheres around the central site).

  Let us discuss the behavior of the model Hamiltonian for the low-energy band
in the Wannier basis.
The kinetic-energy part of the model Hamiltonian (\ref{eqn:TBP}) is a
quadratic function of $\tilde{W}_k^-$. Moreover,
since $\hat{H}$ is periodic in the real space,
its matrix elements
will be diagonal with respect to the momentum $k$ in the reciprocal space:
$\hat{\tilde{\mathfrak{h}}}_k$$=$$\langle \tilde{W}_k^- | \hat{H} | \tilde{W}_k^- \rangle$.
Therefore, the phase ${\rm sgn} [\cos(\theta_k$$-$$\beta)]$
does not contribute to
$\hat{\tilde{\mathfrak{h}}}_k$, and the kinetic-energy part of the model Hamiltonian
\textit{will not depend on the form of the trial orbitals}.
However, for the Coulomb (and exchange) matrix elements, the situation
can be different.
Although the bare Coulomb interaction is also periodic in the real space,
this is a two-particle interaction, and its matrix elements in the reciprocal space
will contain a combination of the four Wannier orbitals:
$\tilde{W}_{k-q}^-\tilde{W}_{k'+q}^-\tilde{W}_{k'}^-\tilde{W}_k^-$.\cite{AGD}
In this case there will be no phase cancelation, and the matrix
elements of the Coulomb interaction will generally depend on the form of
trial orbitals.
Then, one may ask which representation is better? Of course, all are
equivalent and the main question here is whether we would like employ some
additional approximations or not.
A typical approximation is
to retain only the site-diagonal part of
Coulomb (and exchange) matrix elements in the real space,\cite{PRB06}
which may be justified only in the basis of localized WFs
(i.e., for $\beta$$=$$0$ in the considered example).
Then, it is reasonable to expect
the matrix elements between different lattice sites
to be smaller in comparison with the site-diagonal part.
However, the same approximation is no longer valid
for $\beta$$=$$90^\circ$,
where intersite interactions can be of the same order of magnitude as the
site-diagonal ones and, therefore, cannot be dropped.

\subsection{\label{sec:V2O3}Wannier Functions and Model Hamiltonians for V$_2$O$_3$}

\begin{figure}[h!]
\begin{center}
\resizebox{12cm}{!}{\includegraphics{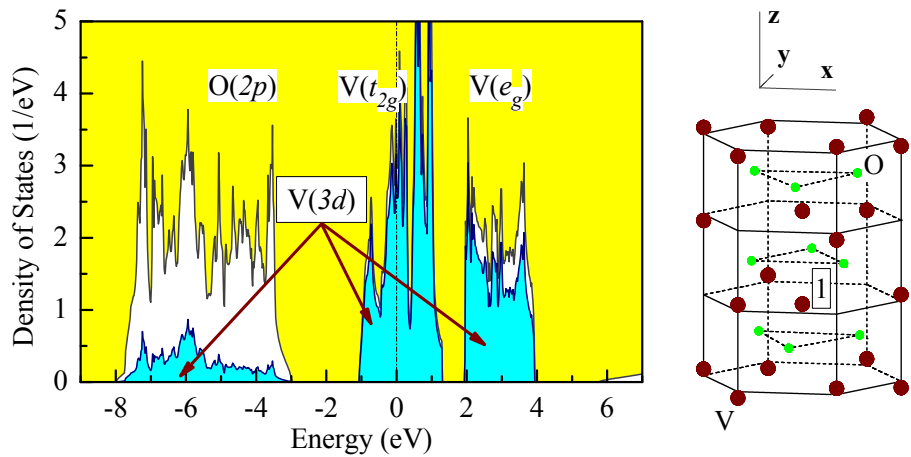}}
\end{center}
\caption{\label{fig.V2O3strucDOS}
(Color online)
Left panel:
Total and partial densities of states of V$_2$O$_3$
in the local-density approximation.
The shaded area shows the contributions of the V($3d$)-states.
Other symbols show the positions of the main bands.
The Fermi level is at zero energy.
Right panel:
the crystal structure of V$_2$O$_3$.
}
\end{figure}

  Now let us illustrate how the same procedure works for more realistic
systems.
For these purposes we pick up rather typical
(and, in some sense, canonical) example of
V$_2$O$_3$ (the space group is
$D^6_{3d}$, No. 167 in the International Tables).
The sketch of the crystal structure as well as the positions of the
main bands of V$_2$O$_3$ in LDA are
summarized in Fig.~\ref{fig.V2O3strucDOS}
(more detailed information
about the lattice parameters
and the basis functions used
the LMTO calculations can be found in Ref.~\onlinecite{PRB06}).
Since each V is surrounded by six oxygen sites, which form a
distorted octahedron, the V($3d$) states are split into the
V($3d$-$t_{2g}$) and V($3d$-$e_g^\sigma$) bands, so that the former ones are
located near in the low-energy part of the spectrum and crossed
by the Fermi level within LDA. The trigonal distortion of the VO$_6$
octahedra will further split the V($3d$-$t_{2g}$) levels into
one-dimensional $a_{1g}$ and two-dimensional $e_g^\pi$ representations.
Since the $e_g^\sigma$ and $e_g^\pi$ orbitals belong to the same representation
of the point symmetry group, they will mix. The magnitude of this
mixing is no longer controlled by the symmetry of the system, and can be
different for different quantities (depending on how these quantities are
constructed from the wavefunctions of V$_2$O$_3$).
In the following, we will call as ``the $e_g^\pi$ orbitals associated with
a site ${\bf R}$''
two eigenstates of the $e_g$ symmetry
obtained from the diagonalization of the density matrix
$\hat{P}_{\bf RR}$$=$$\| \langle \tilde{\chi}_{\bf R}^{\alpha} | \hat{P} |
\tilde{\chi}_{\bf R}^{\alpha'} \rangle \|$,
constructed from the $t_{2g}$ bands, and corresponding to two largest
eigenvalues $\{ \lambda \}$ of $\hat{P}_{\bf RR}$.
Two other eigenstates of the $e_g$ symmetry
for which
$\{ \lambda \}$ are considerably smaller will be called as the
$e_g^\sigma$ orbitals (in realistic calculations for V$_2$O$_3$,
$\lambda$ is about $0.80$ and $0.08$ for the $e_g$ orbitals of the
$\pi$ and $\sigma$ type, respectively).
Thus, the situation is similar to the toy model considered in Sec.~\ref{sec:toy}.

  Then, we take trial orbitals of the $e_g$ symmetry in the
form
$$
| \tilde{\chi}_{\bf R}^{e_g} \rangle =
\cos \beta | \tilde{\phi}_{\bf R}^{e_g^\pi} \rangle +
\sin \beta | \tilde{\phi}_{\bf R}^{e_g^\sigma} \rangle,
$$
and construct WFs for the $t_{2g}$ bands.
In this expressions, $| \tilde{\phi}_{\bf R}^{e_g^\pi} \rangle$ and
$| \tilde{\phi}_{\bf R}^{e_g^\sigma} \rangle$ were obtained from the
diagonalization of the local density matrix, and $\beta$ controls the mixing of the
$e_g$ orbitals of the $\pi$ and $\sigma$ types.
On the contrary, the trial orbital of the $a_{1g}$ symmetry is uniquely
determined by the symmetry
of the system.
Then, according to the
arguments of Sec.~\ref{sec:trial},
$\beta$$=$$0$ should pick up the most localized representation
for WFs. This effect is clearly seen in our calculations,
though there is an important differences from the model considered in Sec.~\ref{sec:toy}.
Note that in the multi-orbital case,
the form of WFs in the reciprocal (${\bf k}$) space is controlled by
a \textit{matrix} $\hat{U}({\bf k})$$=$$\| U_{\alpha \gamma}({\bf k}) \|$
(rather than a single phase factor), which generates
a new set of WFs after the
transformation:\cite{MarzariVanderbilt}
\begin{equation}
| \tilde{W}^\alpha_{\bf k} \rangle \rightarrow
| \tilde{W}^\alpha_{\bf k} \rangle' = \sum_{\gamma} U_{\alpha \gamma}({\bf k})
| \tilde{W}^\gamma_{\bf k} \rangle.
\label{eqn:TransformationMatrix}
\end{equation}
In such a case, the change of $\beta$ will generally induce the change of the
whole matrix $\hat{U}({\bf k})$ [for example, through the additional
orthonormalization transformation (\ref{eqn:orthonormalization})],
which makes some difference from the model
analysis presented in Sec.~\ref{sec:trial}. For instance,
there will be no cancelation of $\hat{U}({\bf k})$ in the
expression for the matrix elements of $\hat{H}$ in the Wannier basis.
Therefore, such matrix elements will generally depend
on the representation $\beta$.
Nevertheless,
we have also observed an abrupt change of WFs in the region of
large $\beta$, in a close analogy with
results of
the model analysis in Fig.~\ref{fig.localizationmodel}.

\begin{figure}[h!]
\begin{center}
\resizebox{12cm}{!}{\includegraphics{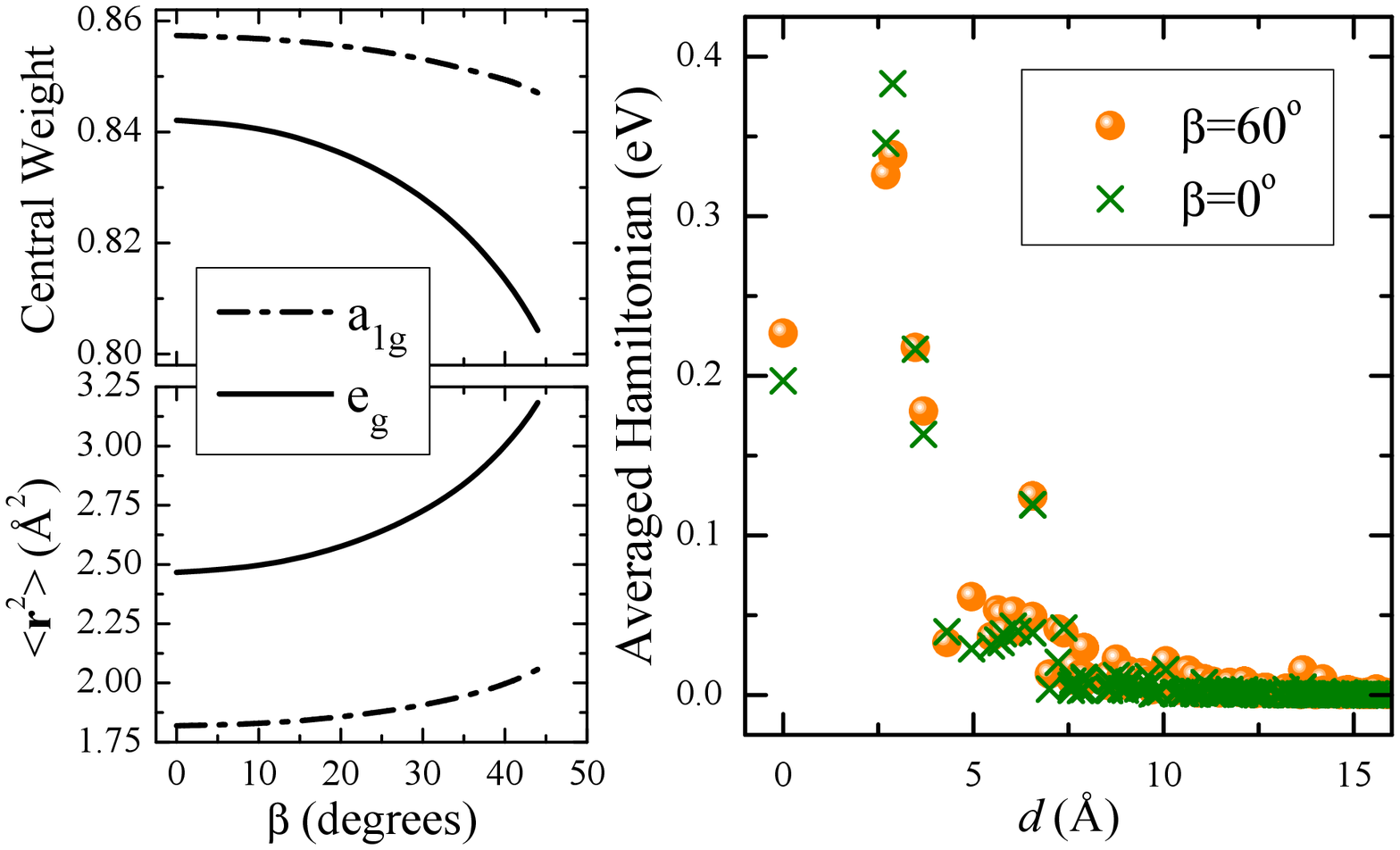}}
\end{center}
\caption{\label{fig.V2O3localization}
(Color online)
The degree of the localization of the
Wannier functions for V$_2$O$_3$ depending on the form of
trial $e_g$ orbitals in the downfolding method. The trial orbitals are
specified by the transformation
$| \tilde{\chi}_{\bf R}^{e_g} \rangle$$=$$
\cos \beta | \tilde{\phi}_{\bf R}^{e_g^\pi} \rangle$$+$$
\sin \beta | \tilde{\phi}_{\bf R}^{e_g^\sigma} \rangle$,
which mixes the $e_g$ orbitals of the $\pi$ and $\sigma$ type. Left panel shows the
weights of the Wannier functions at the central V site (the Central Weight)
and the expectation values of the square of the position operator
versus $\beta$. Right panel shows the distance-dependence of averaged Hamiltonian
$\bar{\mathfrak{h}}_{{\bf RR}'}(d)$$=$$\left( \sum_{\alpha \beta} \tilde{\mathfrak{h}}^{\alpha \beta}_{{\bf RR}'}
\tilde{\mathfrak{h}}^{\beta \alpha}_{{\bf R}'{\bf R}} \right)^{1/2}$
($d$ being the distance between sites ${\bf R}$ and ${\bf R}'$)
for two values of the parameter $\beta$.
}
\end{figure}

   First, let us discuss the results obtained using the downfolding
method for the modified Hamiltonian (\ref{eqn:modifiedH}), which are shown in
Fig.~\ref{fig.V2O3localization}. In practical calculations we set $\epsilon$$=$$10^3$ Ry.
Then, the downfolding method remains stable only up to $\beta$$=$$60^\circ$.
For larger $\beta$, some eigenvalues of $\hat{\cal H}^{rr}$ can fall
into the low-energy part of the spectrum, and the whole procedure becomes
meaningless.
The WFs have been reconstructed from the tight-binding
Hamiltonian (\ref{eqn:TBP}) using the method proposed in Ref.~\onlinecite{PRB06},
which is stable up to $\beta$$=$$45^\circ$.

  As the angle $\beta$ increases,
the weight of the Wannier function of the $e_g$ symmetry at the
central V site decreases from $0.842$ to $0.804$ (i.e., by 5\%), and the expectation value of
${\bf r}^2$, calculated on these WFs, increases from
$2.467$ to $3.183$ \AA$^2$ (i.e., by nearly 30\%). Hence, the WFs
become less localized. Is it a big effect or not? In fact, everything depends
on the considered property.
For example, by assuming for a while that all weight of the Wannier function
is uniformly distributed over the sphere of the radius
$\sqrt{\langle {\bf r}^2 \rangle}$,\cite{PRB94}
we can estimate (very roughly) the
change of the bare Coulomb interaction as 12\%,
which is an appreciable value.

  It is interesting to note that with the increase of $\beta$,
the degree of localization of
the WFs of the
$a_{1g}$ symmetry also decreases,
though the effect is considerably smaller.
This is not surprising, and
the $a_{1g}$ functions should inevitably change because of the additional orthogonalization
of them
to the $e_g$ orbitals.
Note that although the $\tilde{\chi}_{\bf R}^{a_{1g}}$ and $\tilde{\chi}_{\bf R}^{e_g}$
orbitals are orthonormal at the same V site, the matrix elements
$\langle \tilde{\chi}_{\bf R}^{a_{1g}} | \hat{P} |
\tilde{\chi}_{{\bf R}'}^{e_g} \rangle$
between different atomic sites ${\bf R}$ and ${\bf R}'$
are generally nonzero.
Therefore, the WFs of the $a_{1g}$ and $e_g$ symmetry should be
additionally orthonormalized, and this transformation depends on the value of $\beta$.
The same effects leads to the existence of nonvanishing transfer integrals
between $a_{1g}$ and $e_g$ orbitals, which are defined as the
matrix elements of the tight-binding Hamiltonian
$\hat{\tilde{\mathfrak{h}}}_{{\bf RR}'}$ for ${\bf R}$$\ne$${\bf R}'$.

  It is also instructive to consider other possibilities for trial
orbitals, which are found from other principles, and not directly
related with the diagonalization of the density matrix. One possible choice
is the $e_g$ orbitals corresponding to the ideal trigonal environment.
For the site 1 depicted in Fig.~\ref{fig.V2O3strucDOS} such orbitals
have the following form:\cite{Terakura}
$(1/\sqrt{3})(|{\rm yz} \rangle$$+$$\sqrt{2}|{\rm xy}\rangle)$ and
$(1/\sqrt{3})(|{\rm zx} \rangle$$-$$\sqrt{2}|{\rm x}^2$$-$${\rm y}^2\rangle)$.
Another possibility is the so-called ``crystal-field orbitals'' obtained
from the diagonalization of the site-diagonal part of a more general
$5$$\times$$5$ tight-binding Hamiltonian, constructed in the basis of
all V($3d$) orbitals.\cite{PRB06}
Both constructions yield less localized WFs of the
$e_g$ symmetry in comparison with the ones obtained from the
diagonalization of the local density matrix.
Nevertheless, the difference is very small (for example,
$\langle {\bf r}^2 \rangle$$=$ $2.474$ and 2.470 \AA$^2$ for the ideal
trigonal orbitals and the crystal-field orbitals, respectively),
suggesting that in the case of V$_2$O$_3$, a good starting point for the
construction of WFs can be deduced from some other
(for example, geometrical) consideration.
Unfortunately, this rule cannot be applied
equally well for other compounds. For example,
one clear exception is the
distorted perovskite oxides, where the
parameters of the tight-binding Hamiltonian
in the real space are extremely sensitive to the choice
of the local coordinate frame, which is not unique and,
therefore, the construction of the trial orbitals
by means of the diagonalization of the local density matrix
seems to be the most reliable approach.\cite{condmat06}

\begin{figure}[h!]
\begin{center}
\resizebox{9cm}{!}{\includegraphics{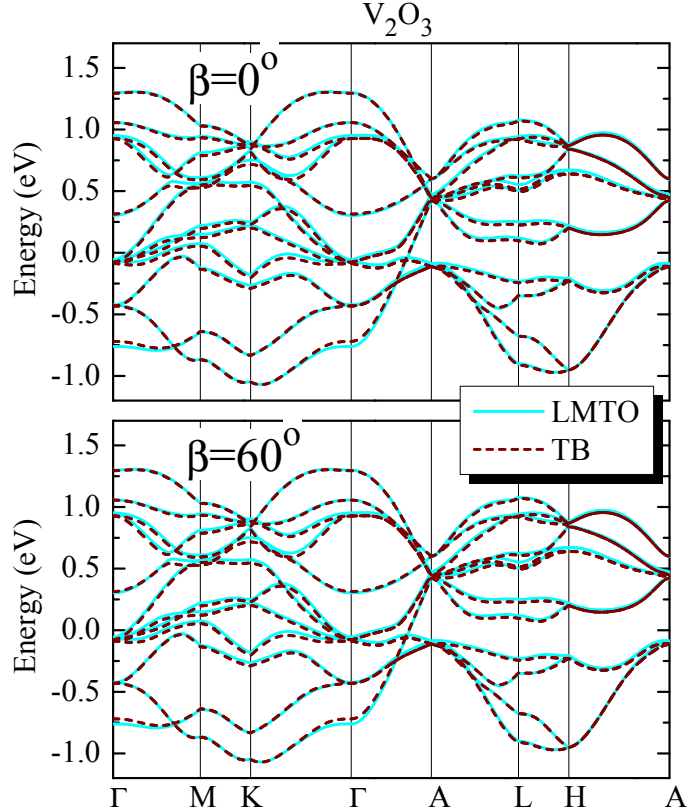}}
\end{center}
\caption{\label{fig.V2O3ek}
(Color online)
LDA energy bands for V$_2$O$_3$ obtained in LMTO calculations and
after the tight-binding (TB) parametrization using the downfolding method
for two values of the parameter $\beta$.
Notations of the high-symmetry points of the Brillouin zone are taken from
Ref.~\protect\onlinecite{BradlayCracknell}.
}
\end{figure}

  In the multi-orbital case,
the degree of localization of the WFs is
related with the distribution of the transfer integrals,
and it is reasonable to expect that
less localized WFs will produce longer-range
interactions in the real space. This is clearly seen in the behavior of
averaged transfer integrals in Fig.~\ref{fig.V2O3localization}.
If for $\beta$$=$$0^\circ$ all nonvanishing transfer integrals
are confined within the radius $d$$\leq$$7.5$ \AA (including 16 coordination
spheres of the V atoms),
those for $\beta$$=$$60^\circ$ have a much longer tail spreading up to
$d$$=$$15$ \AA.
At the same time, the energy dispersion obtained after the diagonalization
of the tight-binding Hamiltonian in the reciprocal space,
$\hat{\tilde{\mathfrak{h}}}_{\bf k}$$=$$\sum_{\bf R} e^{-i{\bf k}({\bf R}-{\bf R}')}
\hat{\tilde{\mathfrak{h}}}_{{\bf RR}'}$, is practically identical for
$\beta$$=$ $0$ and $60^\circ$,
and well reproduce the behavior of the original LMTO bands (Fig.~\ref{fig.V2O3ek}).

\begin{figure}[h!]
\begin{center}
\resizebox{15cm}{!}{\includegraphics{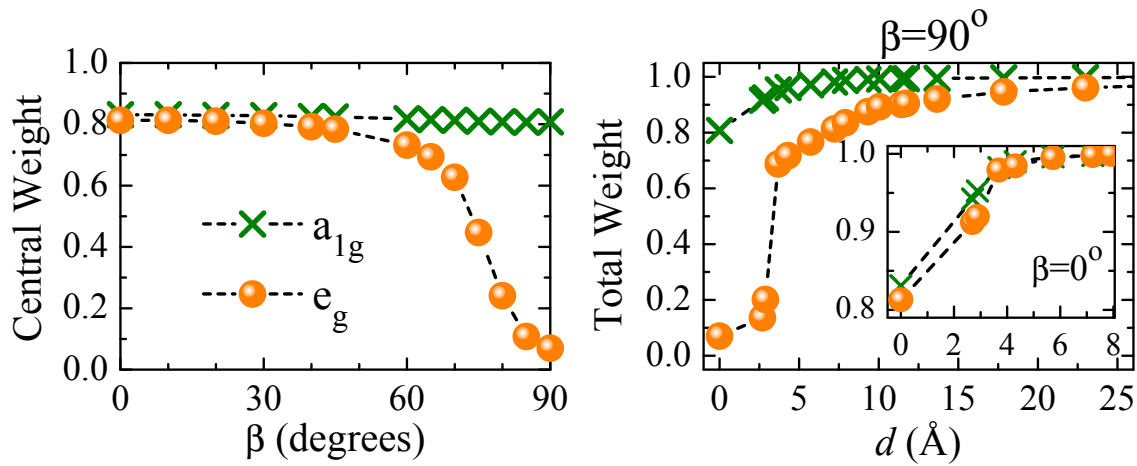}}
\end{center}
\caption{\label{fig.V2O3PO}
(Color online)
The degree of the localization of the
Wannier functions for V$_2$O$_3$ depending on the form of
trial $e_g$ orbitals in the projector-operator method. The trial orbitals are
specified by the transformation
$| \tilde{\chi}_{\bf R}^{e_g} \rangle$$=$$
\cos \beta | \tilde{\phi}_{\bf R}^{e_g^\pi} \rangle$$+$$
\sin \beta | \tilde{\phi}_{\bf R}^{e_g^\sigma} \rangle$,
which mixes the $e_g$ states of the $\pi$ and $\sigma$ type.
Left panel shows the weight of the Wannier functions at the central V site
(the Central Weight)
versus $\beta$. Right panel shows the convergence of the total weight of the
Wannier functions in the real space for $\beta$$=$$90^\circ$.
The inset shows the amplified data for $\beta$$=$$0^\circ$.
}
\end{figure}

  Finally, let us discuss the results of the projector-operator
method. They are summarized in Fig.~\ref{fig.V2O3PO}, showing the weight
of WFs at the central V site as a function of $\beta$ and the convergence of
the total weight of WFs in the real space for two types
of trial orbitals, which are again specified by the parameter $\beta$.
When the angle $\beta$ is not particularly large so that the
downfolding method is applicable, these methods are
identical, as it follows from the rigorous proof in Sec.~\ref{sec:downfoldingasprojector}.
For example, the behavior of the
central weight in Fig.~\ref{fig.V2O3PO} is practically identical
(within the numerical accuracy, which is caused by slightly different details of calculations)
to that obtained for the downfolding method
(Fig.~\ref{fig.V2O3localization}).
However, the projector-operator method,
which can be supplemented practically with any type of the trial orbitals,
allows us to consider more extreme scenarios.
For example, what may happen if our guess about the
form of trial orbitals was totally wrong?
This situation is nicely illustrated by results of our calculations for
$\beta$$=$$90^\circ$ (i.e., by choosing the trial $e_g$ orbital in an orthogonal
subspace to that, which maximizes local part of the density matrix). In this case
the Wannier function simply ``blows up'', and its total weight does not fully converge even
within $d$$\sim$$20$~\AA~around the central V site.\cite{comment.2}
Thus, the situation is very similar to the model analysis in Sec.~\ref{sec:toy}
and apparently related with the discontinuity of the transformation matrix
(\ref{eqn:TransformationMatrix}) in
the reciprocal space.

\begin{center}
\section{\label{sec:summary}Summary and Conclusions}
\end{center}

  We have considered the problem of construction of the Wannier functions in the first-principles
electronic structure calculations starting from localized atomic orbitals.
There is a number of methods, which are used for these purposes. We have
argued that some of them can be unified and formulated in an absolutely exact
and identical way. We have demonstrated this idea for two quite a popular nowadays branches of methods:
the so-called projector-operator method
(considered, for example, in Refs.~\onlinecite{WeiKu}, \onlinecite{Anisimov2005} , and \onlinecite{Streltsov})
and the
downfolding method (considered in Refs.~\onlinecite{PRB04}, \onlinecite{PRB06}, and \onlinecite{condmat06}).
We believe that such an analysis will help to resolve a number of controversies
existing in this area. Of course, identical methods should produce identical results,
and we have shown that the root of the differences between different applications
is related with the choice of the trial orbitals, which are used
in order to
generate the WFs in the projector-operator method or the basis
functions for the
construction of the
effective tight-binding Hamiltonian in the downfolding method.
We believe that it would be very important in the nearest future to make a
similar analysis and generalizations for
the order-$N$ muffin-tin orbital (NMTO) method.\cite{Pavarini,Pavarini2}
However, it is beyond our present abilities.

  The trial orbitals effectively control the localization of WFs.
To this end, we have argued that an optimal choice for the trial orbitals
can be obtained by
maximizing the site-diagonal
part of the density matrix, constructed from the bands of interest.
Our numerical calculations show that such a construction already
provide a good degree of localization for WFs.
Of course, depending on the purposes, one can further optimize the WFs
(for example, by minimizing $\langle {\bf r}^2 \rangle$ or any other
quantity).\cite{MarzariVanderbilt,Nakamura}
However, as a starting point in this procedure one can always use
the WFs generated by minimizing the site-diagonal part
of the density matrix. This is simple and very efficient procedure.
We have also shown that some deviations from this principle
and selection of the trial orbitals in an arbitrary form may lead to
a substantial delocalization of WFs
in the real space, which is related with
discontinuity of their phase in the reciprocal space.

\begin{acknowledgments}
The work of I.~V. Solovyev has been partially supported by Grant-in-Aids
for Scientific Research in Priority Area ``Anomalous Quantum Materials''
from the Ministry of Education, Culture, Sports, Science and Technology of Japan.
Z.~V. Pchelkina and V.~I. Anisimov acknowledge the support by
Russian Foundation for Basic Research under
the grants
RFFI-04-02-16096 and RFFI-06-02-81017.
Z.~V. Pchelkina also acknowledges the support from the
Research Council of President of the Russian Federation
(grant No. NSH-4192. 2006.2),
Russian Science Support
Foundation, Dynasty Foundation, and International Center for
Fundamental Physics in Moscow.
\end{acknowledgments}


\begin{thebibliography}{10}
\bibitem{comment.1}
Note that the matrices $\hat{P}_{\bf R}$ and $\hat{P}_{\bf R}^{1/2}$
have the same set of eigenvectors.
Moreover, if $\hat{P}_{\bf R}^{1/2}$ is positively
defined (that is specified by the choice of the phase factor
in the definition $\hat{P}_{\bf R}$$=$$\hat{P}_{\bf R}^{1/2}$$\times$$\hat{P}_{\bf R}^{1/2}$),
these eigenvectors will be arranged in the same order (i.e., of decreasing eigenvalues).
Therefore,
the additional orthonormalization of WFs (\ref{eqn:orthonormalization})
should not change our conclusion.
\end{thebibliography}
\end{document}